\documentclass[letter]{aa}  

\usepackage{graphicx}
\usepackage{txfonts}
\usepackage{lscape}
\usepackage{hyperref}

\begin{document} 

\title{Hints of auroral and magnetospheric polarized radio emission from the scallop-shell star 2MASS J05082729$-$2101444}
\authorrunning{S. Kaur et al.}
\titlerunning{J0508$-$21 polarized radio emission}
\author{Simranpreet Kaur \inst{1,2}\thanks{E-mail: kaur@ice.csic.es}, Daniele Vigan\`o\inst{1,2,3}, V\'ictor J.S. B\'ejar\inst{4,5}, \'Alvaro S\'anchez Monge\inst{1,2}, \`Oscar Morata\inst{1},  Devojyoti Kansabanik\inst{6,7}, Josep Miquel Girart\inst{1,2}, Juan Carlos Morales\inst{1,2}, Guillem Anglada-Escud\'e\inst{1,2}, Felipe Murgas\inst{4,5}, Yutong Shan\inst{8,9}, Ekaterina Ilin\inst{10}, Miguel P\'erez-Torres\inst{11,12,13}, Mar\'ia Rosa Zapatero Osorio\inst{14}, Pedro J. Amado\inst{11}, Jos\'e A. Caballero\inst{14}, Fabio Del Sordo\inst{1,2,15}, Enric Palle\inst{4,5}, Andreas Quirrenbach\inst{16}, Ansgar Reiners\inst{9}, Ignasi Ribas\inst{1,2}
}

\institute{Institut de Ci\`encies de I'Espai (ICE-CSIC), Campus UAB, Carrer de Can Magrans s/n, 08193 Cerdanyola del Vallès, Barcelona, Catalonia, Spain
\and
Institut d’Estudis Espacials de Catalunya (IEEC), 08860 Castelldefels, Barcelona, Catalonia, Spain
\and
Institute of Applied Computing \& Community Code (IAC3), University of the Balearic Islands, Palma, 07122, Spain
\and
Instituto de Astrofísica de Canarias (IAC), 38205 La Laguna, Tenerife, Spain
\and
Departamento de Astrofísica, Universidad de La Laguna (ULL), 38206 La Laguna, Tenerife, Spain
\and
Cooperative Programs for the Advancement of Earth System Science, Univ. Corporation for Atmospheric Research, Boulder, USA
\and
The Johns Hopkins University Applied Physics Laboratory, 11101 Johns Hopkins Road, Laurel, MD 20723, USA
\and
Centre for Planetary Habitability, Department of Geosciences, University of Oslo, Sem Saelands vei 2b, 0315 Oslo, Norway 
\and
Institut f\"ur Astrophysik, Georg-August-Universit\"at, Friedrich-Hund-Platz 1, 37077 G\"ottingen, Germany
\and
ASTRON, Netherlands Institute for Radio Astronomy, Oude Hoogeveensedijk 4, Dwingeloo, 7991 PD, The Netherlands
\and
Instituto de Astrofísica de Andalucía (IAA-CSIC), Glorieta de la Astronomía s/n, E-18008, Granada, Spain
\and
Center for Astroparticles and High Energy Physics (CAPA), Universidad de Zaragoza, E-50009 Zaragoza, Spain
\and
School of Sciences, European University Cyprus, Diogenes street, Engomi, 1516 Nicosia, Cyprus
\and
Centro de Astrobiolog\'ia, CSIC-INTA, Camino Bajo del Castillo s/n, 28692 Villanueva de la Cañada, Madrid, Spain
\and
INAF, Osservatorio Astrofisico di Catania, via Santa Sofia, 78 Catania, Italy
\and
Landessternwarte, Zentrum für Astronomie der Universität Heidelberg, Königstuhl 12, 69117 Heidelberg, Germany}
  
\date{Received 30 August 2024 / Accepted 25 October 2024}

\abstract
{Scallop-shell stars, a recently discovered class of young M dwarfs, show complex optical light curves that are characterized by periodic dips as well as other features that are stable over tens to hundreds of rotation cycles.  The origin of these features is not well-understood. 2MASS J05082729$-$2101444 is a $\sim$25 Myr old scallop-shell star that was identified using TESS data; it has a photometric period of 6.73\,h that has been attributed to rotation. Of the $\sim$50 recently confirmed scallop-shell stars, it is one of the few detected at radio frequencies between 1 and 8 GHz. We observed this rare system with the upgraded Giant Meterwave Radio Telescope at 575--720 MHz, covering 88\% of the photometric period in each of the two observations scheduled almost a month apart in 2023. We detected approximately millijansky  emission from the target in both epochs, with a significant circular polarization fraction: $|V/I|\sim$20--50\%. The 3.5-minute phase-folded light curves show unique variability in circular polarization. We detected an approximately hour-long helicity reversal during both epochs, and the reversals had similar amplitudes, lengths, and (possibly) occured at similar phases. These results suggest two emission components: The first is a persistent, moderately polarized component possibly ascribable to gyro-synchrotron emission driven by centrifugal breakout events.\ The second is a highly polarized, short burst-like component that is likely due to an electron cyclotron maser (ECM); it is indicative of auroral emission and is potentially responsible for the helicity reversal. To explain this, we discuss the different origins of the plasma responsible for the radio emission, including the possibility that the occulting material is acting as a plasma source. Future coordinated multifrequency radio and optical observations can further constrain the underlying scenario, as well as the magnetic geometry of the system, if we assume an ECM-like auroral emission.}


\maketitle
%

\section{Introduction}\label{sec:intro}

Scallop-shell stars \citep{stauffer17}, also referred to as complex periodic variables \citep{bouma24} or complex rotators \citep{zhan19, gunther22}, are a recently discovered subclass of young ($\sim$5--150 Myr), low-mass stars (mostly M dwarfs) that show clearly periodic, complex optical light curves with chromatic $\sim$5--10\% dips, along with additional features and/or modulations. The rotational periods of scallop-shell stars typically range between 0.2 and 2 days \citep{bouma24}, and the shape of the light curve slowly evolves in time and is maintained over timescales of tens to hundreds of rotation cycles. However, in certain cases, the dips undergo abrupt changes within a single cycle \citep{stauffer17,popinchalk23}. 

The observed optical light curves of a scallop-shell star cannot be adequately explained by traditional models involving only starspots and/or planetary transits. Although starspots can reproduce the chromaticity of the observed dips, they fail to replicate the characteristic sharp features, though they can account for some of the additional features and modulations \citep{stauffer17,zhan19,koen21}. While they do produce similar V-shaped dips, grazing planetary transits cannot account for the mid-term variability and the frequently observed erratic phase shifts. The most convincing scenarios \citep{bouma24} involve the presence of (almost) corotating material around the star, which could be either gas from the star trapped in huge prominences \citep{cameron89, waugh22} or opaque dust-like material from the debris disk or an evaporating rocky planet, similar to the so-far three known cases: KIC 1255b, KOI 2700b, and K2--22b (see \citealt{vanlieshout18} and references within).

In the proposed scenarios, the magnetic field is expected to play a critical role since charged material entrained in the star's magnetic field naturally appears at the corotation radius \citep{bouma24}. Furthermore, the scallop-shell stars bear similarities with classical magnetic rotators that have unusual magnetic topologies, such as magnetic B stars \citep{Townsend}. Therefore, probing the magnetism of these objects may add an important piece to the puzzle. Although the classical Zeeman-Doppler imaging techniques \citep{semel89, piskunov93, kochukhov16} often cannot be used to infer the magnetic topology of very low-mass stars due to the intrinsic faintness of the stars and the relatively low signal-to-noise ratio, a few spectropolarimetric studies have extended the sample to mid-to-late M dwarfs \citep{morin10, Berdyugina_LSR}. However, for many M dwarfs, it is possible to infer their magnetic properties, albeit with non-negligible uncertainties, by modeling the spectral line broadening with radiative transfer calculations \citep{reiners22}. Alternatively, and more directly, radio emission from M dwarfs can trace their magnetic activity and flaring mechanisms \citep{mclean11, bloot24}.

Radio stars are rare, though a few recent studies, based on the cross-matching of different catalogs, are extending the known samples for M dwarfs especially \citep{vedantham20,callingham21,yiu24,driessen24,kao24}\footnote{See also the Sydney Radio Star Catalogue: \url{radiostars.org}}. In particular, \citealt{yiu24} cross-matched sources in the LOw-Frequency ARray (LOFAR) Two-metre Sky Survey (LoTSS and V-LoTSS for Stokes $I$ and $V$, respectively; \citealt{shimwell22,callingham23}) and the Very Large Array (VLA) Sky Survey (VLASS; Stokes $I$ only; \citealt{gordon21, gordon23}) with the \textit{Gaia }database of nearby ($d<$ 100\,pc) sources \citep{gaia21}. They found only a couple dozen stars at LOFAR frequencies (40--120 MHz) and 65 Stokes $I$ sources at 2--4 GHz. Most of these radio-loud stars are M dwarfs, and only a few are confirmed as scallop-shell stars by \cite{bouma24}. In this study we focus on one such radio-loud scallop-shell star, 2MASS J05082729$-$2101444 (hereafter J0508$-$21).


J0508$-$21 was previously detected at the millijansky level in short pointings by the \textit{Karl Jansky} VLA and the Australian SKA Pathfinder Telescope (ASKAP) at gigahertz frequencies (see Appendix~\ref{app:archival}). In this Letter, we present our sub-gigahertz observations of J0508$-$21 using the upgraded Giant Meterwave Radio Telescope (uGMRT). They are the longest dedicated radio observations of any scallop-shell star to date and show very peculiar trends, which are possibly related to the optical dips.

This Letter is organized as follows: Sections \ref{sec:Properties} and \ref{sec:Radio_observations} explain the known properties of the target, the radio observations, and the data reduction. We present our results in Sect. \ref{sec:Results} and conclude with a discussion on the possible scenarios responsible for the observed trend in radio emission in Sect. \ref{sec:discussion}.

\section{Target properties and photometric information} \label{sec:Properties}

J0508$-$21 (RA 05h08m27.30s, Dec $-$21d01m44.40s) is classified as an M5.0\,V dwarf \citep{riaz06}. It lies at a distance of $48.30\pm0.14$\,pc \citep{gaia21} and is likely a member of the $\beta$ Pictoris moving group (but see \citealt{lee24}). The presence of lithium in the spectra is consistent with the system being young \citep{ribas23}, with an estimated age of 10--30 Myr (\citealt{miret20} and references within). The source shows UV and X-ray excess, which is typical of active young M dwarfs. Stellar magnetic fields are expected to be strong (due to the young age and likely fast rotation), but they are so far unconstrained.

The target was recently classified as a scallop-shell star by \cite{bouma24}, who inspected the 2-minute-cadence Transiting Exoplanets Survey Satellite (TESS) data in sectors 5 and 32. A periodogram over the available TESS sectors gives a best-fitting main period of $P=0.280455\pm0.000035$ days (i.e., $\sim$6.73 hours; \citealt{shan24}). 
The TESS light curves \citep{bouma24} have shown photometric modulation and regular dips at a similar phase of rotation since at least 2018, albeit with a slight evolution in morphology. Complementary ongoing photometry campaigns of the source over the years have found the dips to be persistent but with a slow phase shift (for this particular timing solution), evolving in shape after hundreds or  thousands of cycles, which is typical of scallop-shell stars. In this study we took the latest available photometric light curves into account; they were observed with the Las Cumbres Observatory Global Telescope (LCOGT) in March 2023 (see details in Appendix~\ref{App:LCO}). 

\begin{table*}
\caption{uGMRT observations.}
\label{table:1}      
\centering          
\begin{tabular}{c c c c c c c c}      
\hline\hline    

&  & \multicolumn{3}{c}{Stokes $I$} & \multicolumn{3}{c}{Stokes $V$ } \\    
\cline{3-8}
Date &Clean beam &Peak & Integated   & $\sigma_{I}$   & Peak         & Integrated & $\sigma_{V}$ \\
(dd/mm/yy) & ($''\times''$) & (mJy/beam) & (mJy) & (mJy/beam) & (mJy/beam) & (mJy) & (mJy/beam)\\
\hline                    
11/06/23  & $5.75\times4.02$ & $1.094 \pm 0.017$ & $1.076 \pm 0.030$ & 0.014 & $0.160 \pm 0.010$ & $0.183 \pm 0.021$ & 0.009\\  
29/07/23  & $4.96\times4.09$ & $1.299 \pm 0.014$ & $1.217 \pm 0.023$   & 0.013 &$0.200 \pm 0.009$  & $0.188 \pm 0.016$ & 0.008\\

Combined   & $4.99\times3.92$& $1.200 \pm 0.013$ & $1.190 \pm 0.023$ & 0.012 & $0.185 \pm 0.007$  & $0.183 \pm 0.012$ & 0.006\\ 
\hline                  
\end{tabular}
{\tablefoot{The columns give the date, clean beam size, peak, and integrated fluxes of the source and the rms of the self-calibrated clean images, for Stokes $I$ and $V$, for each observation and their combined image. Each observation lasted 7 hours, including overheads. The fluxes were calculated using the task {\tt imxfit} and performing a Gaussian fitting. The errors in the peak and integrated flux represent the statistical standard deviation from the Gaussian fitting.}
}
\end{table*}

\section{Radio observations and data reduction} \label{sec:Radio_observations}

\subsection{uGMRT observation setup}

We observed J0508$-$21 in uGMRT band 4 (550--900\,MHz) for 14 hours in total as part of proposal 44\_045. The data were recorded in the left-handed ($LL$) and right-handed ($RR$) polarization with an integration time of 2.68 seconds. The observations were split into two sessions of 7 hours each, scheduled 1.6 month apart in June and July 2023 (see Table~\ref{table:1}). Including the overheads, each observation allowed us to cover on-source $\sim$88$\%$ of the target's photometric period of $6.73$ hours. In both observations, 3C147 was used as the flux and bandpass calibrator, while J0521$-$207 was used for phase calibration. In each observation, the flux calibrator was observed once at the beginning and once at the end, while the phase calibrator was observed repeatedly after every target scan of 28 minutes.

\begin{figure}
\centering
\includegraphics[width=0.45\textwidth]{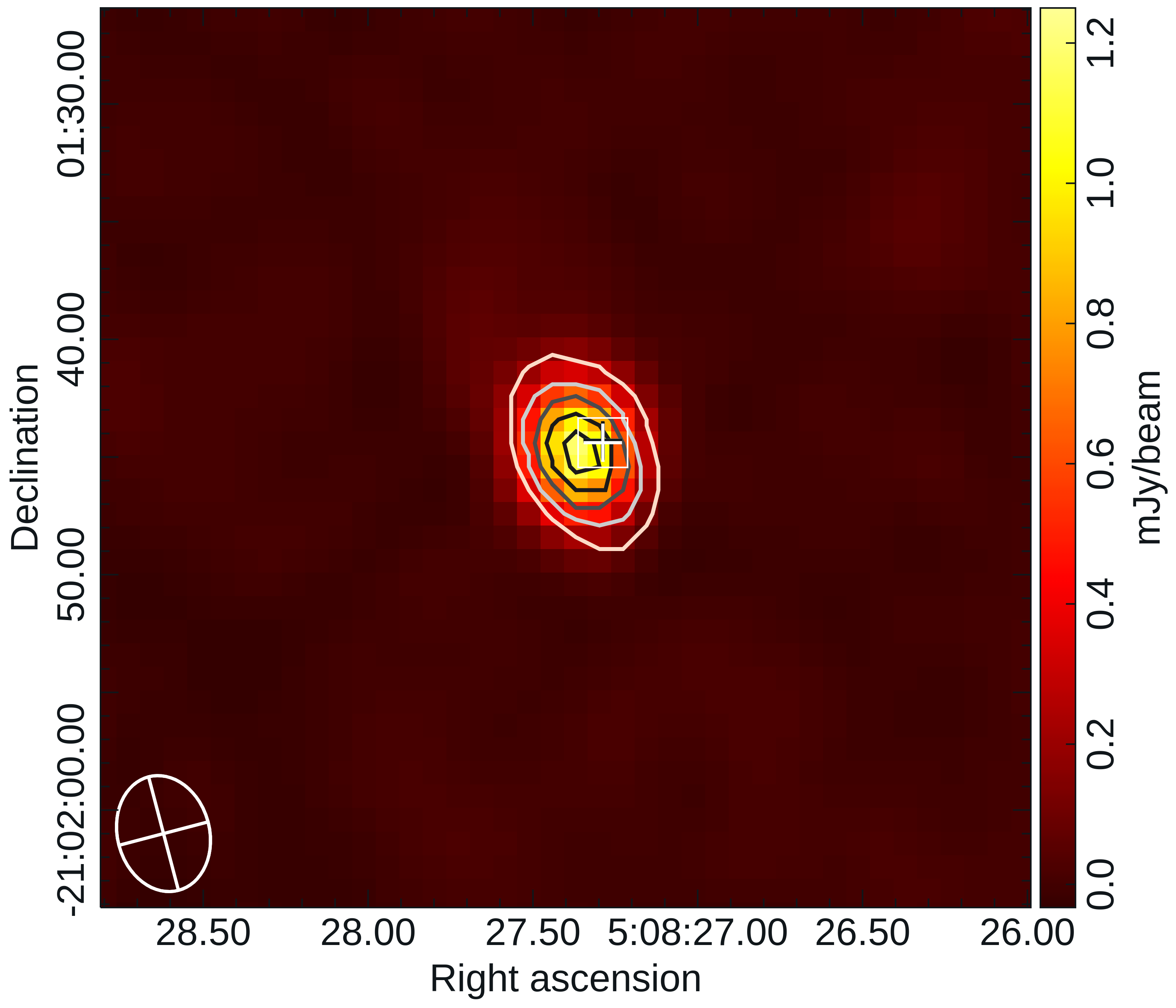}
\caption{Combined map for the June and July uGMRT radio observations centered on J0508$-$21. The color scale represents the Stokes $I$ flux density, while the contours indicate the Stokes $V$ signal, corresponding to levels of 5, 10, 15, 20, and 25 $\sigma^V_{rms}$ (where $\sigma^V_{rms}$ is $\sim$ 6.5~$\mu$Jy/beam). The synthesized beam is shown in the lower-left corner, and the cross marks the position of the target.}
\label{fig:gmrt}
\end{figure}

\subsection{Flagging, calibration, and imaging}

We processed and imaged the dataset using the {\tt CAPTURE} pipeline (CAsa Pipeline-cum-Toolkit for Upgraded Giant Metrewave Radio Telescope data REduction; \citealt{kale21}). This pipeline executes tasks like flagging, calibration, imaging, and self-calibration utilizing Common Astronomy Software Applications (CASA; \citealt{CASA}). After the flagging and standard calibration, we visually inspected the data to identify the remaining bad data. The resulting usable band was 575--720 MHz.

We corrected for the instrumental delay and bandpass using the bandpass calibrator, and applied the solutions to the phase calibrator. We noticed that a residual, non-corrected delay was still present in the data, affecting both datasets. This delay was obvious when plotting the corrected phase versus frequency of the phase calibrator; it appeared as a spectral slope that was particularly steep for the longer baselines, with a large phase difference at the extremes of the frequency band. We used the phase calibrator, which has sufficient flux density ($\sim$4 Jy), to perform an additional delay correction and applied the solutions to all our targets.

The maps were constructed with the CASA task {\tt tclean} by using the W-projection gridding algorithm to correct for wide-field non-coplanar baseline effects \citep{cornwell08} and employing a Briggs weighting scheme with {\tt robust=0} to have a good compromise between the beam size and sensitivity. We also carried out three rounds of self-calibration using the bright sources in the field to improve the sensitivity in the final images. Since Stokes $V$ and Stokes $I$ signals are both formed from the $RR$ and $LL$ correlation products, with Stokes $I=(RR + LL) / 2$ and Stokes $V=(LL - RR) / 2$ (where the latter definition follows uGMRT standards, which are opposite to the International Astronomical Union (IAU) convention followed by, e.g., VLA; \citealt{das20,chandra23}), we did not perform any additional polarization calibration as it is more important for the cross-hand products. In addition to using {\tt CAPTURE}, we also double-checked the results with an independent, in-house analysis script, which confirmed all the results presented here within the statistical errors.

\begin{figure*}[t]
\centering
\includegraphics[width=.95\textwidth]{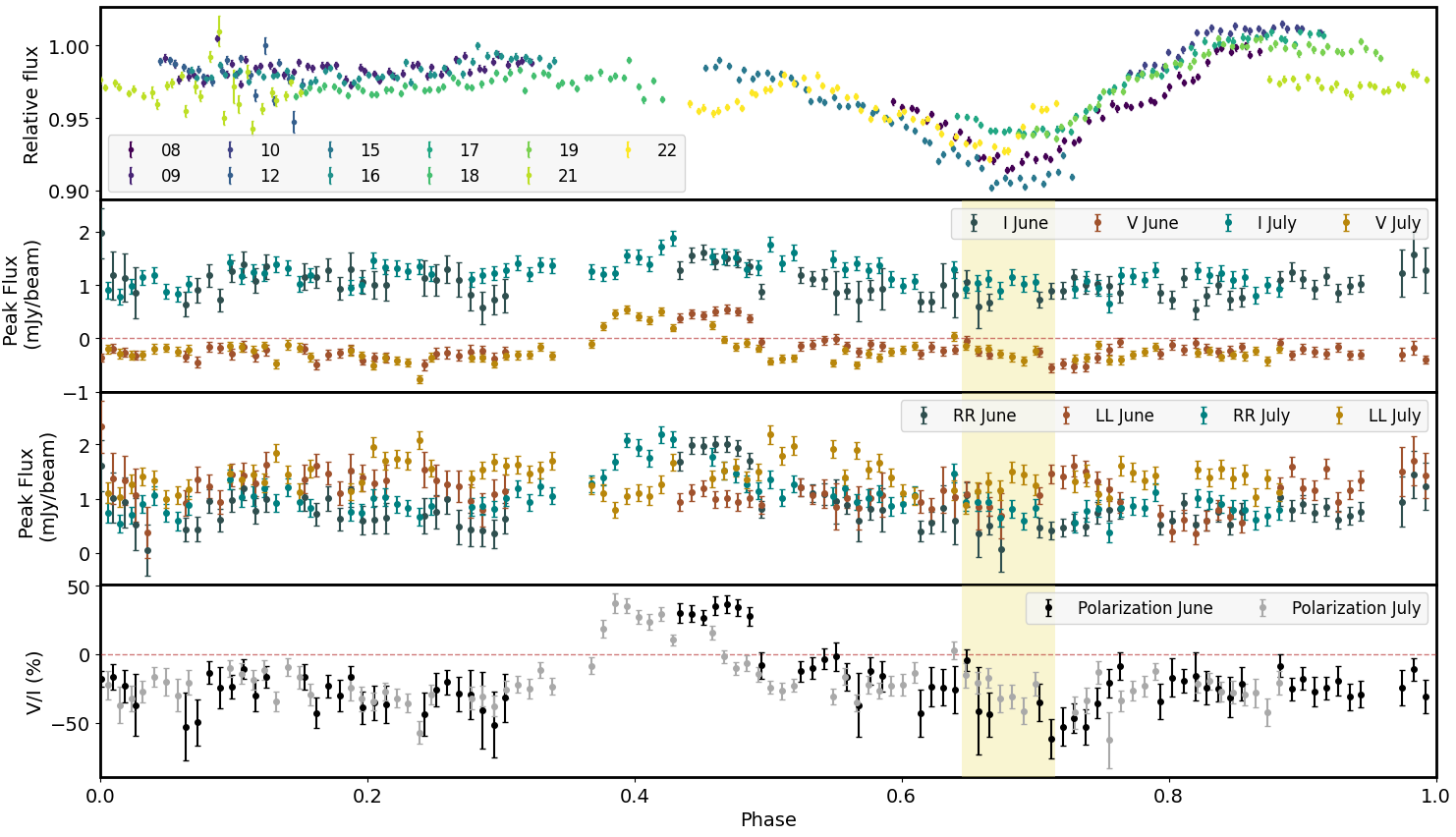}
\caption{Phase-aligned optical and radio light curves of J0508$-$21, assuming P=0.280455 days and taking the beginning of the first target scan in the July uGMRT observation, $T_{0}(JD)$ = 2460154.51215, as the reference phase ($\phi=0$). {\em Top panel:} Optical $i$-band fluxes from the 40 cm camera of LCOGT, with colors marking different days of the March 2023 campaign (see Appendix~\ref{App:LCO} for details). {\em Second panel:} Radio light curves (Stokes $I$ in green, Stokes $V$ in brown) of the 2023 uGMRT observations in June (darker shades) and July (lighter shades). Each horizontal point represents the peak flux integrated over 3.5 minutes, in band 4 (575-720 MHz). The vertical error bars represent the statistical error $1~\sigma_{rms}$ for each time binning. {\em Third panel:} $RR$ (brown) and $LL$ (green) components for June and July observations, with the same color-code as the top panel. {\em Bottom panel:} Circular polarization fraction $V$/$I$, where we have excluded a few points with a very large ($\sim$100$\%$) relative error. The yellow-shaded phase range indicates the expected location of the main optical dip at its minimum, as extrapolated from the LCOGT observations (Bejar et. al. in prep) and with the assumed timing solution. Its width indicates the phase error of the dip center, obtained by propagating the estimated error on the period $\sigma_P \sim 0.000035$ days.} 
\label{fig:radio_0508}
\end{figure*}

\subsection{Flux and light curve extraction}

We used the CASA task {\tt imfit} to measure the flux densities, sizes, and positions of the Stokes $I$ and $V$ components from the clean maps. To investigate the variability in flux over time for each session, we generated clean maps for short, 3.5-minute intervals for both Stokes $I$ and $V$, and analyzed the statistics of each map using the task {\tt imstat}. We also double-checked the results from the visibility plane, as explained in Appendix~\ref{App:DS}.

We looked for possible variations in the real part of the amplitude for $LL$ and $RR$ correlators by integrating the data for 3.5 minutes and over all spectral windows, and derived the corresponding variation in Stokes $I$ and Stokes $V,$ respectively. Along with studying the behavior of the flux with time, we also investigated the spectral behavior of the target (see Appendix~\ref{App:DS}).

\section{Results} \label{sec:Results}

\subsection{Images and fluxes}

We detected a clear Stokes $I$ and $V$ signal at the position of the target in both uGMRT observations. Figure~\ref{fig:gmrt} shows the resulting map, combining both observations.  
For each individual observation, we obtain a peak brightness of 1.07 and 1.25\,mJy/beam (S/Ns of $\sim$74 and 93) in Stokes $I$ signal and 0.15 and 0.18\,mJy/beam in Stokes $V$ signal (S/Ns of $\sim$16 and 22) for the June and July observations, respectively. The average Stokes $I$ flux is $\sim$10--15\% larger in June than July, while the Stokes $V$ fluxes are statistically similar, with a corresponding average circular polarization fraction of $\sim$15$\%$ but very high variability (see Sect. \ref{sec:variability}).
Table \ref{table:1} summarizes the flux statistics, obtained from the Gaussian fitting of the clean maps with the CASA task {\tt{imfit}}. The similar peak and integrated flux values are compatible with a point source, that is, no extended emission is detected beyond the beam size (projected physical size of $\sim200\times240$~au). 

We also looked into the archival observations over the years and found the target to be consistently radio-loud, with variable emission at approximately millijansky levels, up to 8\,GHz at least (see Appendix~\ref{app:archival} for details).

\subsection{Time variability and circular polarization reversal}\label{sec:variability}

Since our observations cover $\sim$88\% of the photometric period of the target twice, we were able to study how the flux intensity and circular polarization vary in time. Figure \ref{fig:radio_0508} shows the abovementioned LCOGT optical light curves (top) and our new uGMRT radio light curves: Stokes $I$ and $V$ (second panel), $LL$ and $RR$ components (third panel), and polarization fraction $V$/$I$ (bottom panel). These curves were produced with a time binning of 3.5 minutes, a reasonable compromise between the time resolution and relative errors at each point.

The light curves are shown as a function of phase, folding the observations with the TESS best-fitting period (Sect. \ref{sec:Properties}). In both observations, the $RR$ and $LL$ components are always nonzero, with the $LL$ component moderately larger most of the time. However, the $RR$ component has an important increase around phase $\phi$$\sim$0.35--0.5 in both observations. This excess translates to a temporary helicity reversal in Stokes $V$ signal that lasted less than one hour and led the expected phase of the photometry dip (i.e., extrapolated from LCOGT data) by $\delta\phi$$\sim$0.15--0.25.

Overall, the Stokes $V$ flux varies between about $-$0.5 and 0.5\,mJy/beam, being close to zero only briefly when changing signs. The bottom panel of Fig. \ref{fig:radio_0508} shows that the resulting circular polarization fraction varies between about $-$60\% and $+$50\%, and it lies between about $-$20\% and $-$30\% most of the time. Due to the helicity reversal, the instantaneous Stokes $V$ signal and the corresponding polarization fraction are almost always higher than the $\sim$15$\%$ obtained from each full (or combined) epoch integration mentioned above. The emission is extended across the entire frequency band at all times, and we cannot infer a reliable spectral slope due to the limited bandwidth (see Appendix~\ref{App:DS} for dynamic spectra).

Still, we find some differences between the two epochs. In June, the $RR$ excess is not accompanied by any $LL$ variation, while in July, the $LL$ starts increasing when the $RR$ component decreases, possibly causing a higher overall flux excess (Stokes $I$, light green in the second panel). Although the June observations started in the middle of the main $RR$ excess, it seems that the peaks in June and July are slightly mismatched in phase, with a difference of $\sim$0.05 that is not entirely ascribable to the period uncertainty (which propagates to a $\sim$0.02 phase uncertainty over 1.6 months). We also note that in both epochs, approximately hour-long smaller variations happen at different phases. Nevertheless, there are striking similarities between the two light curves, in particular in terms of the intensity, duration, and phase range of the Stokes $V$ reversal.

\section{Discussion and summary} \label{sec:discussion}

We investigated the short-term variability of radio emission from the scallop-shell star J0508$-$21 based on two 7-hour-long observations at sub-gigahertz frequencies using the uGMRT. We confirm the source is a persistent radio-loud star at millijansky levels over a broad frequency range (at least $\sim$0.5--8\,GHz), a rare behavior observed within a class of objects that contains only $\sim$50 known good quality sources within 150 pc \citep{bouma24}. The most striking results are the persistent circular polarization in both epochs and its hour-long helicity reversal, which is possibly modulated by the 6.73-hour optical period. 

\subsection{Emission mechanisms explaining the radio light curve}

Our preferred interpretation of the light curve is the presence of two components, similar to what was inferred for, for example, the ultra-cool dwarf LSR J1835+3259 \citep{kao23,climent23} or T Tauri S \citep{johnston03,loinard07}. The first component appears to be persistent, quasi-steady, moderately circularly polarized, and possibly ascribable to magnetospheric gyro-synchrotron emission (e.g., \citealt{berger01,burgasser05,hallinan15}). This component might be supported by a mechanism similar to the one operating in the centrifugal magnetospheres of the early magnetic B-type stars, which explains the observed common behavior of their incoherent radio emission \citep{leto21, shultz22}, a commonality that seems to hold in general for gyro-synchrotron radio emission from the ordered magnetospheres of ultra-cool dwarfs and Jupiter \citep{leto21}. In this scenario, magnetic reconnection driven by a continuous centrifugal breakout helps  accelerate electrons, resulting in gyro-synchrotron emission that varies with the stellar rotation period \citep{Owocki_2022}.

The second component is a coherent, almost purely $RR$-polarized hour-long emission, possibly electron cyclotron maser (ECM) emission, detected only during a small range of phases (width of $\sim$0.15). ECM requires the presence of a magnetic field with strong intensity, $B$, and an anisotropic distribution of electrons (see Appendix~\ref{app:illustration} for a brief summary of ECM ingredients), and it results in a beamed emission at the gyro-frequency $\nu$ [MHz]$\simeq$2.8\,B[G] or, at most, its second harmonic \citep{zarka07}. Since $B$ has spatial variations, an ECM source usually shows a broadband spectrum with a cutoff at the gyro-frequency that corresponds to the largest value of $B$ in the emitting region.

The uGMRT light curve shown here seems unique as it displays a helicity reversal of circularly polarized emission that lasts shorter compared to Stokes $V$ reversals found in literature.\  Such reversals have been seen at very different timescales and lower polarization fractions  ($\lesssim 5\%$) in T Tauri \citep{skinner94,johnston03} or as sinusoidal modulations of fully polarized emission coming from two magnetic poles \citep{mclean11,williams17}. We tentatively attribute the absence of a sinusoidal variation in our observations to the ECM emission originating from only one magnetic pole, although we keep in mind that the emission driven by ECM strongly depends on the observed frequency band \citep{das21}, the magnetic field strength, and the geometry (the locations of the emission, observer, and the star magnetic and rotation axes).

Another way to identify the underlying mechanism is to look at the phenomenological classes of radio bursts proposed by \cite{bloot24} for M dwarfs (mainly AU Mic but also, e.g., UV Ceti and YZ Ceti; \citealt{villadsen19,zic19,bastian22,pineda23}). We tentatively classify the short-term $RR$ excess seen here as Type C (slow sweeps in frequency, 1--3 hour long), D (fast sweeps, 0.5--1 hour long), or F (irregular bursts, 1--6 hour long). The limited bandwidth of our observations (less than 200\,MHz of usable band) does not allow a clearer identification, but we can state that all of these classes are an expression of the ECM, which is often associated with auroral processes (e.g., \citealt{hallinan07,pineda23}). An alternative is coherent plasma emission \citep{dulk85} associated with flares, though this phenomenon has not been confirmed in any star other than the Sun. However, plasma emission produces intrinsically lower fluxes since it is less beamed. Moreover, given that TESS data for J0508$-$21 show $\sim$20 flares randomly distributed across the phase, it seems highly unlikely that two flares, occurring 1.6 months apart, would coincidentally align at the same phase to produce such similar and bright radio emission.

\subsection{Hints of auroral emission?}

The statistically relevant collection of AU Mic bursts over 250 hours of observations \citep{bloot24} shows a clustering of $RR$ and $LL$ short-lived excesses around two different phase ranges. In this context, the $RR$ excesses for J0508$-$21 could arise from beamed ECM emission, visible only during a given phase range and coming from one magnetic pole. The apparent small phase mismatch of $\sim 0.05$ between the June and July $RR$ excess can be explained by the intrinsic irregularity of the emission (see the dispersion in phase from one episode to another in Fig. 7 of \citealt{bloot24}), seen also for the Jupiter--Io system \citep{zarka07}. 



For a scallop-shell star, a natural source of plasma to power the ECM could be the same opaque, approximately corotating material responsible for the optical dip. Of the two scallop-shell star scenarios outlined in Sect. \ref{sec:intro}, the plasma-trapped-in-prominence scenario is slightly disfavored, because the optical dips are quite regular in phase over the years (see the end of Sect. 5.8 of \citealt{bouma24}), and one would expect prominences to disappear or move much faster. In the second scenario (i.e., ionized dust at the corotating radius), the regularity of the dips over the years suggests the presence of an evaporating object on a stable orbit (rather than small clumps, which might live on shorter timescales of a few months) \citep{sanderson23}. In this case, J0508$-$21 would host a much younger and maybe more extreme example of the few rocky evaporating planets discovered by \textit{Kepler} \citep{brogi12,rappaport12,rappaport14,schlawin18,vanlieshout18}; this planet would be responsible for both the optical dips and for providing the plasma necessary for ECM emission. In this case, the magnetic interaction between the stellar magnetic field and the material then would be akin to a Jupiter--Io-like interaction (Io decametric emission), and one would expect a systematic phase lag between the optical dip and the right-handed circularly polarized radio emission  ($\sim 0.25$ in our two radio observations and the closest optical dip).

However, an issue for both scenarios (prominences or evaporating planet) is that, in order to dissipate power and accelerate particles, there needs to be relative motion between the plasma and the stellar magnetic field (i.e., a deviation from the corotation of the material). In this sense, an alternative, possibly more plausible scenario is that the occulting material only acts as an indirect plasma source filling a plasma torus, similar to the Jupiter--Io and Saturn--Enceladus systems, and, more generally, the stellar magnetosphere. Such plasma could trigger ECM emission that is detectable at certain phases for an observer, similarly to the non-Io decametric, hectometric, and kilometric radiation related to the main auroral oval of Jupiter \citep{zarka01, zarka04}. In this case, one would not expect any relation between the optical dip and the ECM emission. A slight variation of this is that the plasma is provided by the star itself, possibly like for fast-rotating brown dwarfs \citep{kao18}, and the occulting material plays no role. The peculiar light curve would in this case be caused by the complex magnetic field topology that M dwarfs are known to possess \citep{morin10}, with the $RR$ excess corresponding to a very specific geometrical or magnetic configuration, and there would be no connection with the optical dip. In Appendix \ref{app:illustration}, we show a qualitative picture of these scenarios.

While with the available radio data we cannot determine which scenario is most likely, they add suggestive pieces to the puzzle of scallop-shell stars. To draw more substantial conclusions, we would need:
(i) a radio follow-up over a broad range of frequencies; this would allow us to determine if the Stokes $V$ emission has a cutoff frequency compatible with what is expected from an ECM ($\nu_{cut}$ of $\simeq 2.8~B$[G] MHz), which would give us information about the local magnetic field, $B$;
(ii) repeated, long sub-gigahertz observations that cover the entire period, possibly simultaneously or contemporaneously with optical observations, which would allow us confirm and to better characterize the intriguing behavior seen in these two observations;
(iii) a comprehensive characterization of the system architecture based on radial velocity, spectroscopic, and imaging observations; and (iv) similar monitoring of the other radio-loud scallop-shell stars from the small \cite{bouma24} sample.

\begin{acknowledgements}
SK carried out this work within the framework of the doctoral program in Physics of the Universitat Aut\`onoma de Barcelona. SK, OM, and DV are supported by the European Research Council (ERC) under the European Union’s Horizon 2020 research and innovation programme (ERC Starting Grant "IMAGINE" No. 948582). We acknowledge financial support from the Agencia Estatal de Investigaci\'on (AEI/10.13039/501100011033) of the Ministerio de Ciencia e Innovaci\'on and the ERDF ``A way of making Europe'' through projects
  PID2022-137241NB-C4[1:4],    
  PID2021-125627OB-C31,        
  PID2020-117710GB-100,         
  PID2020-117404GB-C21,         
  RYC2021-032892-I,             
and the Centre of Excellence ``Severo Ochoa''/``Mar\'ia de Maeztu'' awards to the Institut de Ci\`encies de l'Espai (CEX2020-001058-M), Instituto de Astrof\'isica de Canarias (CEX2019-000920-S), and Instituto de Astrof\'isica de Andaluc\'ia (CEX2021-001131-S). 
We thank the staff of the GMRT that made these observations possible. GMRT is run by the National Centre for Radio Astrophysics of the Tata Institute of Fundamental Research, India. This work makes use of observations from the LCOGT network. We also thank the referee for the valuable and constructive feedback.
\end{acknowledgements}

\bibliographystyle{aa}
\bibliography{output}

\onecolumn
\begin{appendix}

\section{Archival observations}\label{app:archival}

\begin{figure}[h]
\centering
\includegraphics[width=0.75\textwidth]{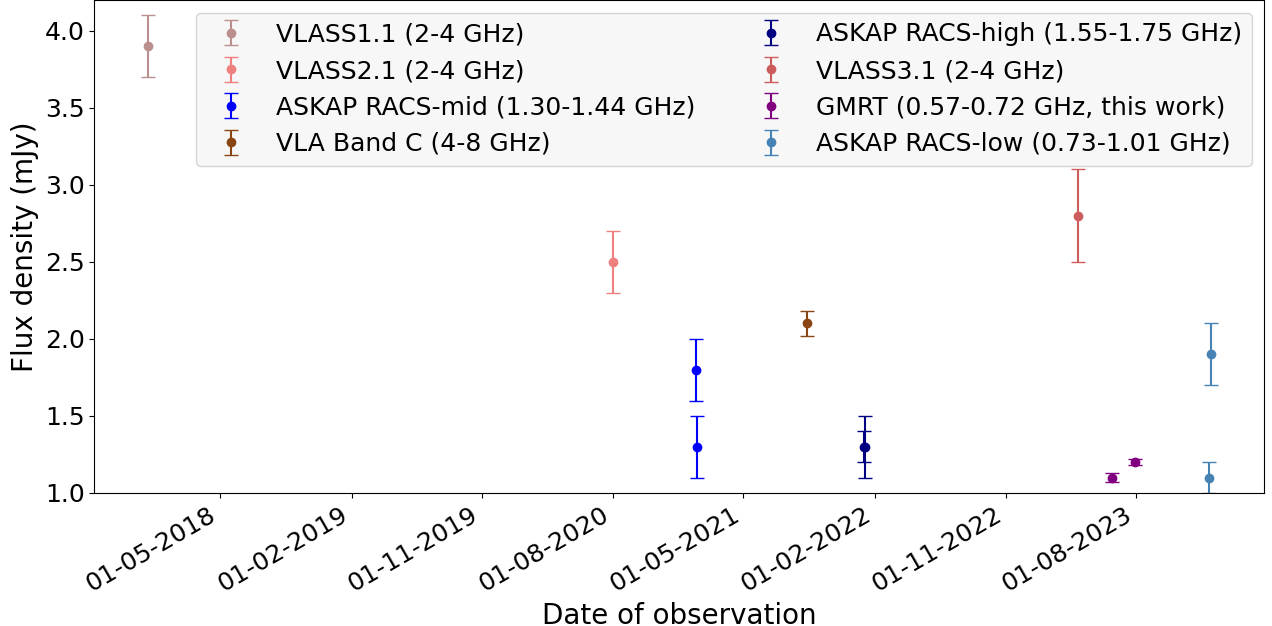}
\caption{All known Stokes $I$ measurements (flux densities and related 1$\sigma$ errors) for J0508$-$21 in different bands, as indicated by the instruments and representative frequencies in the legend.}
\label{fig:overall_lc}
\end{figure}

In Fig.~\ref{fig:overall_lc} we present the flux density of the source obtained from archival observations over the years. They include: (i) the VLASS survey (S band, 2--4\,GHz), with a Stokes $I$ flux density of $3.9 \pm 0.2$ mJy, $2.5 \pm 0.2$ mJy and $2.8 \pm 0.3$ mJy in epochs 1.1, 2.1 and 3.1, respectively; (ii) the relatively narrowband all-sky Rapid ASKAP Continuum Survey (\citealt{driessen24}) at 0.887 GHz (RACS-low; \citealt{mcconnell20,hale21}), 1.367 GHz (RACS-mid; \citealt{duchesne23}), and 1.655 GHz (RACS-high), with $\sim$1--2\,mJy Stokes $I$ detections and shallow upper limits on Stokes $V$ (e.g., $\sigma_V\sim 0.22$ mJy in 1.6 GHz in the RACS-High survey pointings); and (iii) one single, short (10 minutes on-source) observation with VLA band C (4--8\,GHz, project 21A-349). We performed the analysis of the latter dataset, using the available calibrated visibilities and the standard CASA VLA pipeline: we found a Stokes $I$ detection, also reported by \cite{shabazz21}, with a flux density of $2.09 \pm 0.06$\,mJy. There is no Stokes $V$ detection ($\sigma^V_{rms}\sim$12~$\mu$Jy/beam) in the 10-minute integrated image; however, when splitting the observation into five 2-minutes bins, we found a $\sim$6$\sigma$ Stokes $V$ signal in the last 4 minutes, peaking at $85~\mu$Jy/beam, unreported so far. Unfortunately, most of these data do not provide Stokes $V$ information. Since the flux values measured in VLASS data are systematically higher than the others, one can tentatively claim that the spectra peaks around band S. However, the data points are too sparse in time to have a clear idea of the broadband spectral distribution, considering the evident (but poorly characterized) flux variability. However, importantly, these observations confirm that J0508$-$21 is consistently radio-loud between 0.55 and 8\,GHz at least.

\section{LCOGT optical observations}\label{App:LCO}

We observed J0508$-$21 over 11 nights between 2023 March 8 and 22 using the $i$ band of the 40~cm telescopes of LCOGT \citep{Brown2013} at the Siding Spring Observatory. We obtained 50 individual exposures of 120~s, giving a total observing time of about 2 hours per epoch. The 40~cm telescopes have a 3k$\times$2k SBIG CCD camera with a pixel scale of 0.571\,arcsec, providing a field of view of 29.2$\times$19.5\,arcmin. 
Weather conditions were clear most of the days and the average seeing varies from 2.0 to 6.5\,arcsec.
Raw data were processed using the {\tt BANZAI} pipeline \citep{McCully2018}, which includes bad pixel, bias, dark and flat field corrections for each individual night. 
We performed aperture photometry of the star and several reference stars with similar brightness, and obtained the relative differential photometry 
using AstroImageJ \citep{Collins2017}, adopting an optimized aperture of 10 pixels (5.7\,arcsec). The overall on-going photometric and spectroscopic monitoring of the source will be presented in detail in a dedicated upcoming paper (B\'ejar et al. in prep). In the top panel Fig.~\ref{fig:radio_0508} we show the corresponding light curves.

\section{Dynamic spectra}\label{App:DS}

\begin{figure*}[h]
\centering
\includegraphics[width=.49\textwidth, height = 6.5cm]{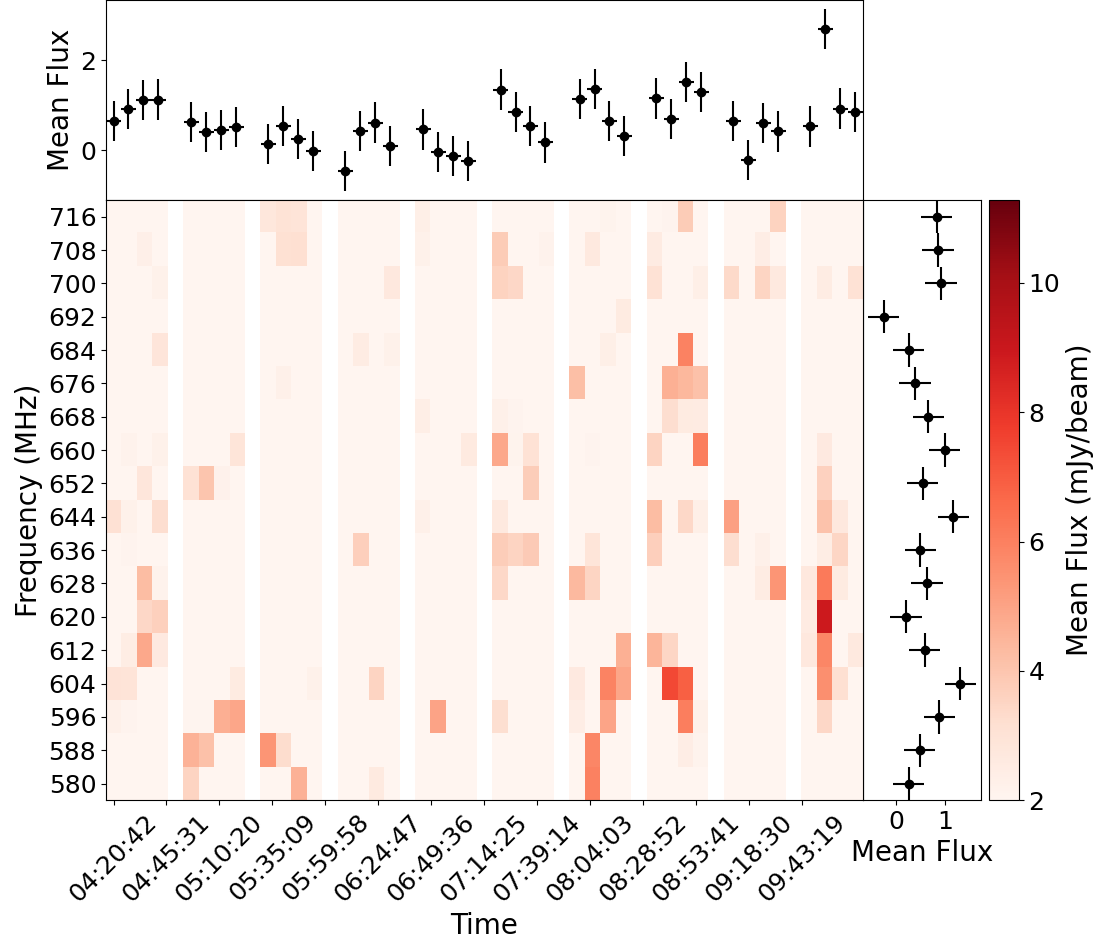}
\includegraphics[width=.49\textwidth, height = 6.5cm]{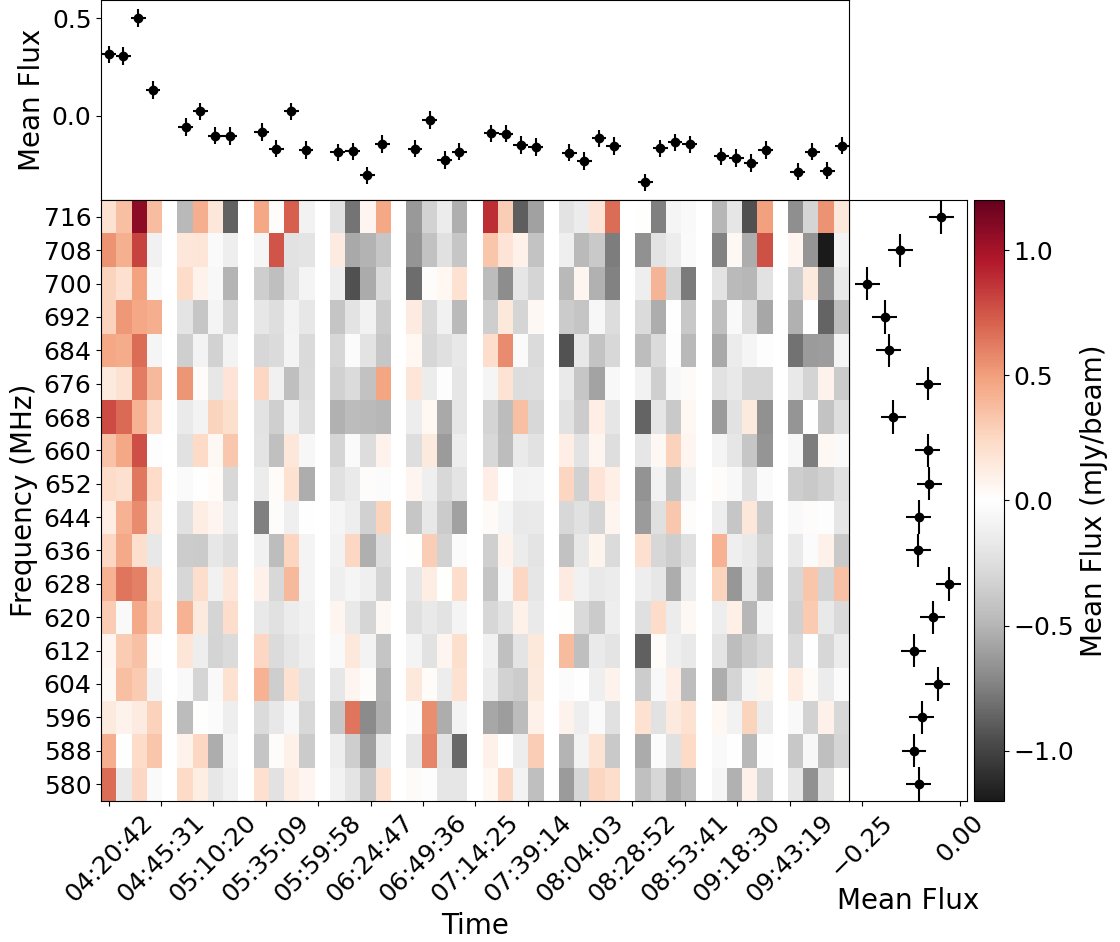}
\includegraphics[width=.49\textwidth, height = 6.5cm]{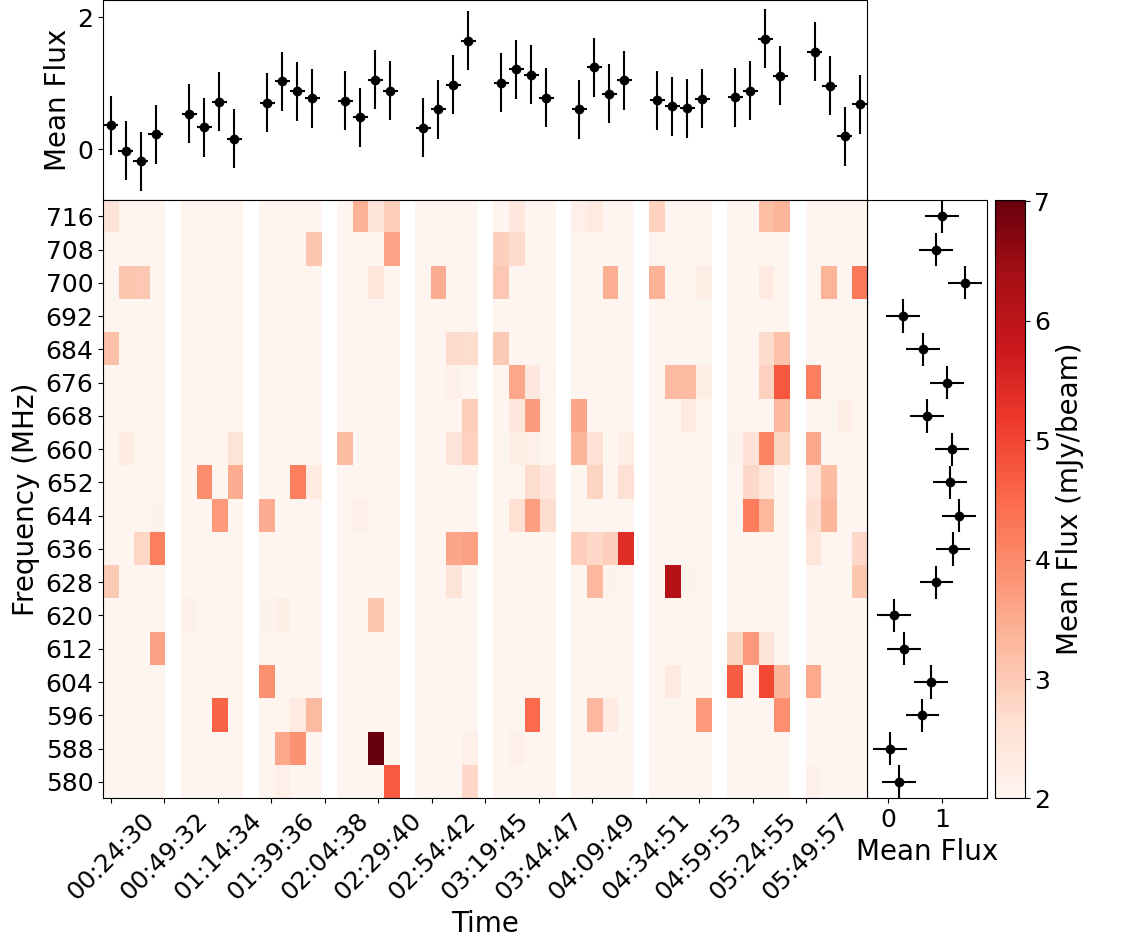}
\includegraphics[width=.49\textwidth, height = 6.5cm]{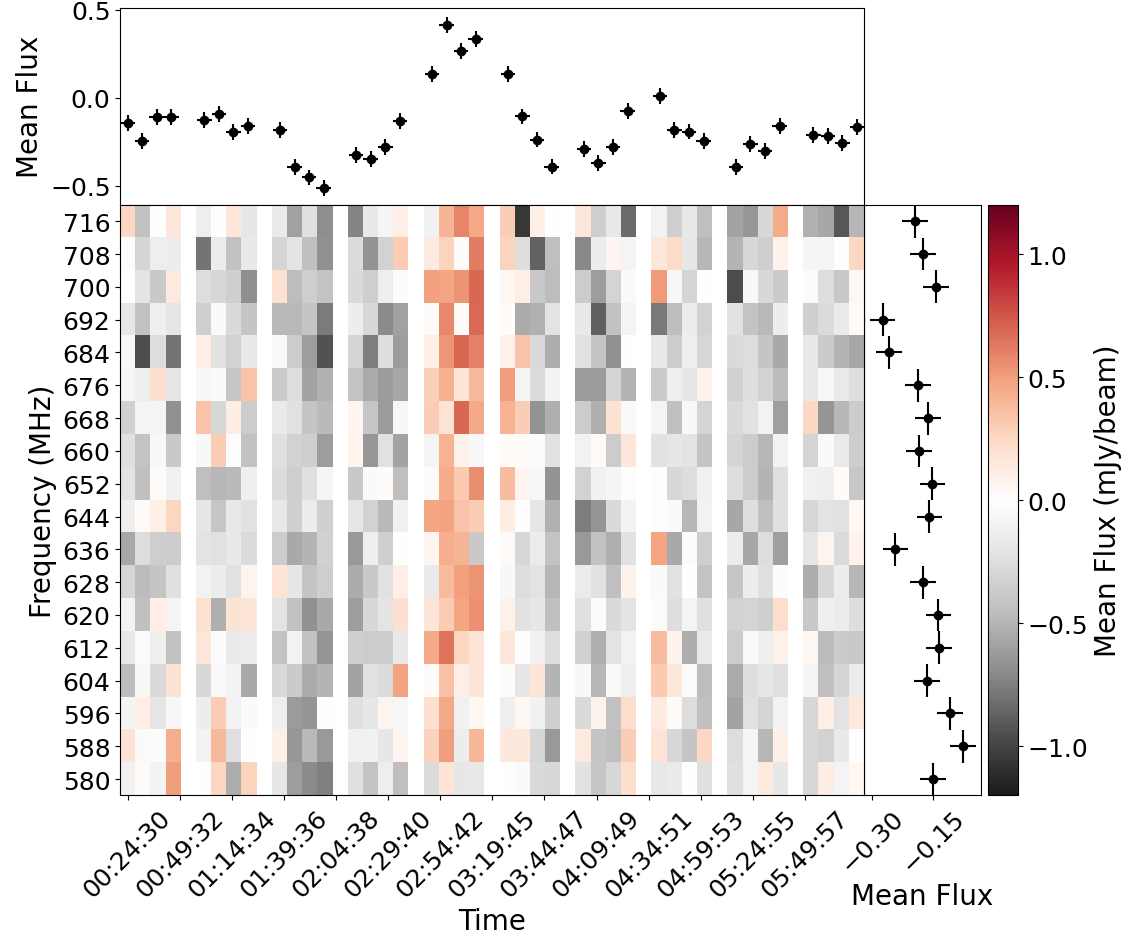}
\caption{Dynamic spectrum of Stokes $I$ (left) and Stokes $V$ (right) emission for June (top) and July (bottom) observations. Each bin corresponds to an integration time of 7 minutes averaged over 40 channels ($\sim$8 MHz), a good compromise between resolution and S/N per bin. The average rms for each bin in Stokes $I$ is $\sim 2$mJy/beam, while it is $\sim0.32$ mJy/beam for Stokes $V$. The narrow white gaps mark the time spent on the phase calibrator during the observation. In each panel, the top and right subsets show the time series and the average spectra, obtained by integrating the dynamic spectrum in frequency and time, respectively.}
\label{fig:DS_V}
\end{figure*}

Figure~\ref{fig:DS_V} shows the dynamic spectrum of a region encompassing the source for both observations (top and bottom, respectively), for Stokes $I$ (left) and $V$ (right), in the usable band, 575--720 MHz. To obtain these dynamic spectra, we used the CASA tasks {\tt specflux} and {\tt imstat} after generating the cube images with {\tt tclean}, integrated over 7 minutes. In each plot, we show the time-averaged spectra in the right subset, the frequency-integrated light curves in the top subset. The latter are compatible with the ones shown in Fig.~\ref{fig:radio_0508}, but the time resolution is twice as coarser. From these plots, the emission appears to be broadband at all times. The Stokes $V$ reversal is clear at all frequencies, with possible but unclear signs of lasting slightly longer at higher frequencies. The average spectra do not show a clear trend neither in Stokes $I$ nor $V$: the bandwidth and the S/N are not statistically significant to have a reliable estimate of the slope. In Stokes $I$, the plot shows hints of a few narrowband, drifting features, but again, the S/N is not high enough for a sound claim.

We have also double checked the dynamic spectra and the time series shown in Fig.~\ref{fig:radio_0508} from the visibility plane. For that, we used the CASA task {\tt visstat} to calculate flux statistics in the visibility plane. The variability analysis from the visibility plane was done after subtracting bright ($>3\sigma$) background sources from the visibilities using the CASA task {\tt uvsub}. To do this, we created a model of the bright sources from the self-calibrated $RR$ and $LL$ maps with {\tt tclean}, excluding the region around our source, and then subtracted these models from the corrected visibilities. This ensured that the flux statistics for our target were less affected by any bright background emissions, although some features can persist, especially if the background sources are variable. Note that obtaining such plots from the visibility plane is a common choice for practical reasons (as it is much faster to obtain), but it introduces many more uncertainties and technical caveats compared to using the cube images, mostly due to the contamination from the imperfect cleaning of the brightest sources within the primary beam (few tens of arcminutes). In our case, this is especially true for Stokes $I$, since there are several sources with similar or larger flux intensities (see, e.g., Sect. 2 of \citealt{bloot24} for a summary of the technical challenges in using dynamic spectra from the visibilities). Despite these caveats, the dynamic spectra from the visibility plane are qualitatively similar to the ones obtained from the cube images shown here.

\section{Speculative scenarios of auroral emission}\label{app:illustration}

\begin{figure*}[h]
\centering
\includegraphics[width=0.3\textwidth]{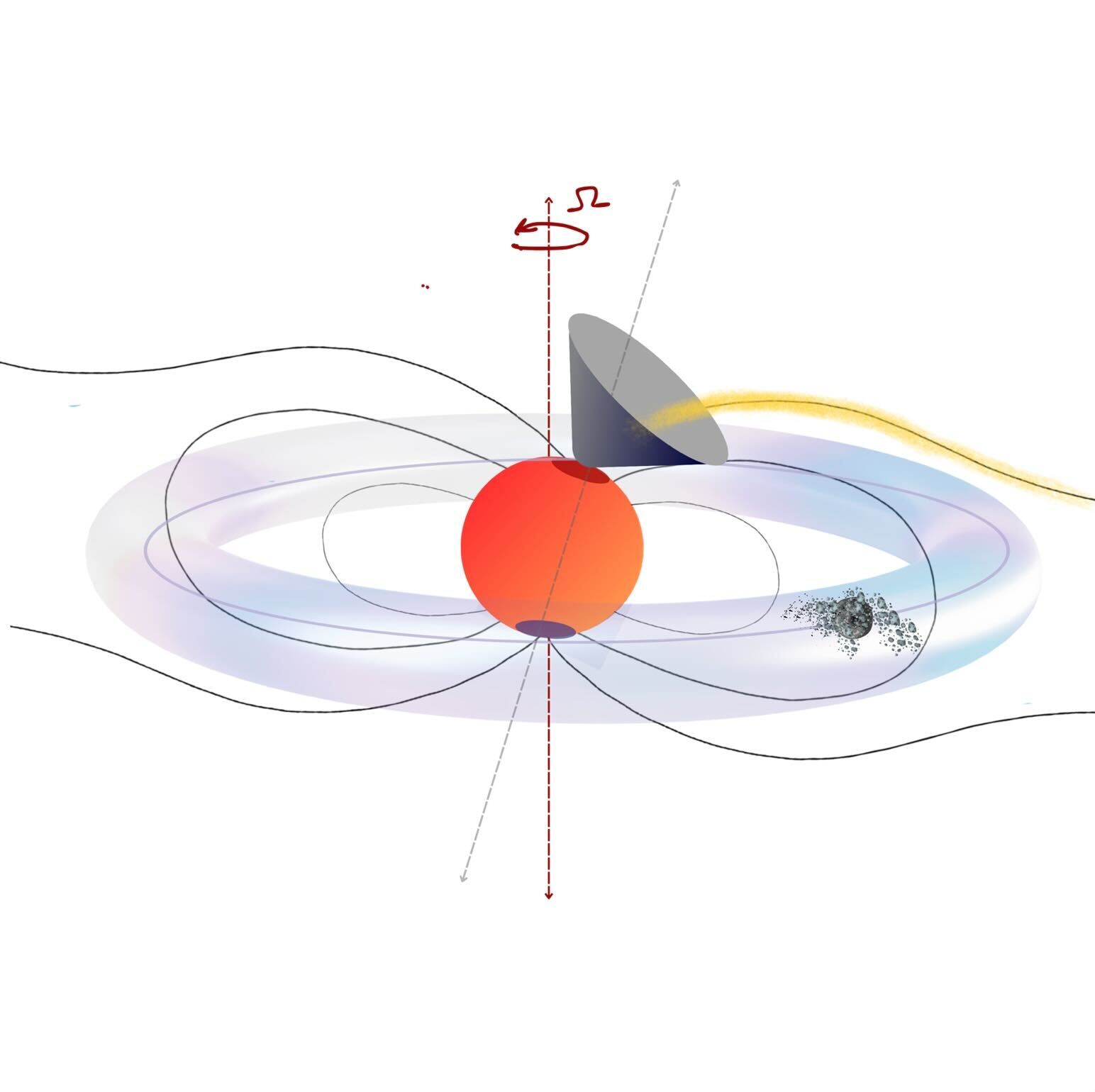}
\includegraphics[width=0.3\textwidth]{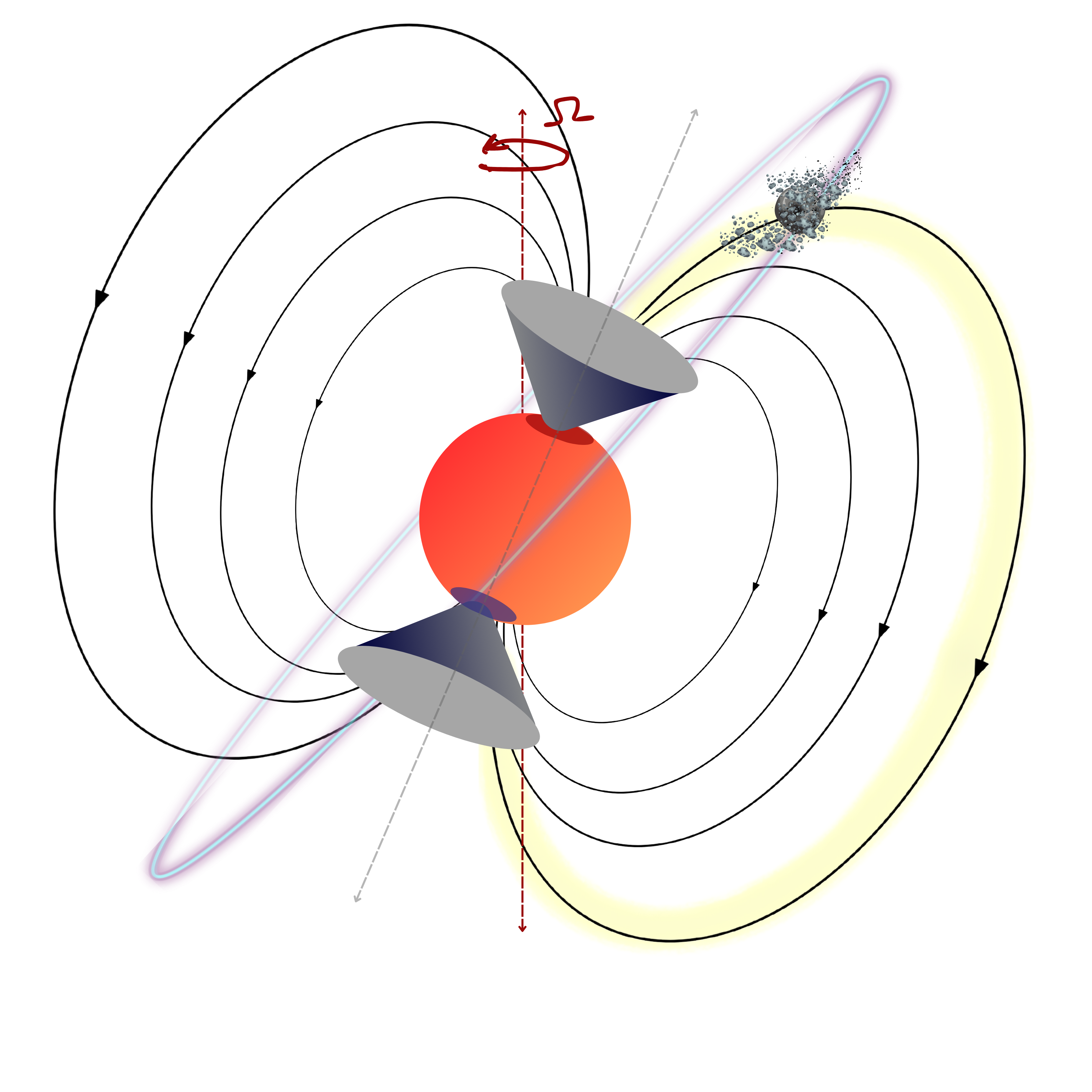}
\includegraphics[width=0.3\textwidth]{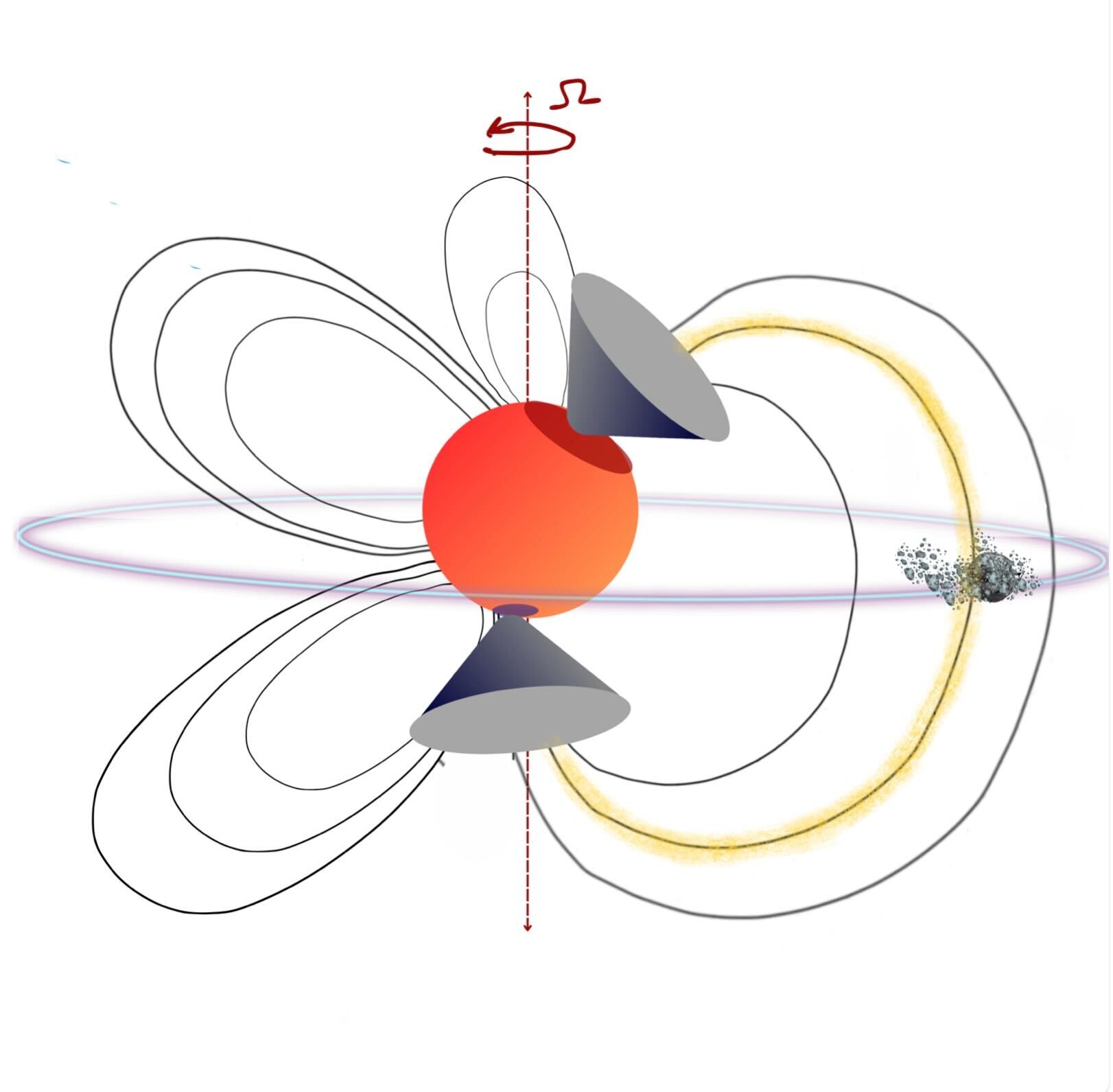}
\caption{ Illustrations of possible scenarios that could trigger the beamed radio ECM emission along a wide cone. The yellow lines qualitatively mark the flux tubes at the footprints from which ECM emits radio-beamed emission, marked by the gray cones. In all cases, the orbit of the occulting material is seen nearly edge-on by the observer, which explains the optical dips. {\em Left:} Non-Io decametric-like emission, for which the occulting object might be only an indirect source of plasma. {\em Center:} Approximately corotating, evaporating planet causing optical dips, with a very inclined orbit. {\em Right:} Same as the center, but for a complex, multipolar stellar magnetic field planetary orbit orthogonal to the stellar rotation axis.}
\label{fig:illustration}
\end{figure*}

In general, ECM arises in the presence of a magnetic field with strong intensity $B$ and a population of energetic electrons that have an anisotropic and unstable energy distribution, like loss-cone or horse-shoe distribution \citep{winglee86, treumann06}. The anisotropy in the electron velocity distribution provides the free energy necessary to amplify electromagnetic waves, leading to intense, highly polarized radio emissions, the frequency of which depends on the local magnetic field $B$, as $\nu$ [MHz]$\simeq$2.8\,B[G], or, at most, its second harmonic \citep{Melrose_dulk_1982,treumann06}. Auroral ECM is emitted along the wall of a hollow cone with a wide-opening angle, thus it should be modulated, with minutes- to hour-long bursts concentrating at specific phases, as clearly seen in AU Mic. In principle, the geometrical model presented in \citep{bloot24} could be used to constrain the geometry of system. However, at contrast with \cite{bloot24}, we only have two datasets (one of which does not cover the entire $RR$ excess), therefore we cannot convincingly constrain the free parameters in the model (viewing angle, magnetic inclination, footprint co-latitude, opening angle and width of the hollow cone, and, in the multipolar case, the size and location of the small-scale field line).

Therefore, here we just give the qualitative representation of the speculative scenarios discussed in the main text, illustrated in Fig.~\ref{fig:illustration}. We show three cases. The first one (left panel) is non-Io decametric-like, where the plasma responsible for the emission activates magnetic field lines close to the magnetic poles. In this case, plasma is unrelated to the occulting object, but could originally come from it or its plasma torus. According to the continuous centrifugal breakout scenario, the centrifugal force acting on the co-rotating ionized material breaks the magnetic field lines at a large distance, where the magnetic field strength is no longer able to trap the co-rotating plasma. This effect locally accelerates non-thermal electrons, which then travel towards the stellar surface, emitting radiation via incoherent gyro-synchrotron and, possibly, coherent ECM, if this electron population also develops an unstable energy distribution.

The other possibilities (central and right panels) is the Io-decametric-like scenario, for which the plasma released by the evaporating object is magnetically connected to the surface. If the occurrence of only $RR$-polarized episodes at a given phase range is confirmed by future observations, we could be seeing the magnetically-induced ECM coming only from one magnetic pole of the star. As a matter of fact, assuming that the optical dip is caused by the same object that induces the radio emission, its orbit has to be edge-on. Since a simple configuration with an orbit orthogonal to the stellar rotation axis and a dipolar field would give rise to ECM with both polarities (see, e.g., the AU Mic radio model by \citealt{bloot24}), which we do not observe, we need a more complex configuration. Qualitatively, seeing only one magnetic pole could be due to either due to (i) a highly non-orthogonal viewing angle $i_*$, which allows us to see the auroral emission from only one magnetic pole, if a magnetic tilted dipole similar to \cite{bloot24} is employed (central panel), or (ii) an orthogonal viewing angle but strongly non-dipolar field lines (right panel), in line with the complex field configurations expected for a young, active M dwarf. In both cases, the ECM is expected to happen on a wide-open cone, so that the radio emission is seen with a certain phase delay compared to the optical dip. Such delay for the Io-Jupiter system is about $\delta\phi$$\sim$0.25 (meaning, we receive the ECM radiation when Io is at quadrature), not far from what we might be observing.

In this evaporating planet scenario, the day-to-day morphological irregularity of optical dips (see the top panel of Fig.~\ref{fig:radio_0508}) and radio light curves could be associated with chaotic processes when material present in the form of an irregular cloud suddenly becomes unstable in orbit after the accumulation of ejected mass from the orbiter and drains through, for example, one of the four Lagrange points of the system close to the planet. Note also that the co-rotating radius is the natural innermost possible orbit and can lead to a synchronization between the stellar spin and the planetary orbit, as seen, for example, in the $\tau$ Boo system \citep{walker08}. On the other side, in order to have non-zero dissipation to radio power, we need to assume a deviation from strict co-rotation, as mentioned in the main text as a major concern.

In any case, at this stage, we cannot support or quantify the likelihood of these scenarios, which are speculative and raise other questions, like the absence of a detected cometary-like tail, seen in some, but not all, evaporating planets, \citealt{vanlieshout18}. These scenarios should be tested using future multiwavelength, coordinated observations.

\end{appendix}

\end{document}